\documentclass[10pt]{article}

\usepackage[T1]{fontenc}
\usepackage{newtxtext, newtxmath}
\usepackage{amsmath}

\usepackage{siunitx}
\DeclareSIUnit\angstrom{\protect \text {Å}}
\usepackage[version=4]{mhchem}

\usepackage{graphicx}
\usepackage[margin=1in]{geometry}
\usepackage{microtype}
\usepackage[super,sort&compress,comma]{natbib}

\usepackage[font=sf, labelfont=bf]{caption}

\usepackage{setspace}
\usepackage[australian]{babel}

\begin{document}

\title{On (not) deriving the entropy of barocaloric phase transitions \\from crystallography and neutron spectroscopy}
\author{Anthony E. Phillips \and Helen C. Walker}

\maketitle

\begin{abstract}
  We review well-known signatures of disorder in crystallographic and inelastic neutron scattering data. We show that these can arise from different types of disorder, corresponding to different values of the system entropy. Correlating the entropy of a material with its atomistic structure and dynamics is in general a difficult problem that requires correlating information between multiple experimental techniques including crystallography, spectroscopy, and calorimetry. These comments are illustrated with particular reference to barocalorics, but are relevant to a broad range of calorics and other disordered crystalline materials.
\end{abstract}

\section{Introduction}
\subsection{Rationale}
Barocaloric materials, which can be switched between low- and high-entropy states by applying pressure, are 
promising candidates to replace vapour-compression refrigerants. 
Until recently, such materials were thought to be rather rare; however, work in the past decade has shown that many, perhaps even most solid-solid phase transitions are in principle barocaloric.\cite{boldrin_fantastic_2021, lloveras_advances_2021} 
The key fundamental challenge facing the field is therefore to understand the entropy changes in known materials well enough to be able to identify, and even design, new materials likely to show phase transitions with large entropy changes.

Such entropy changes are determined by a material's atomistic structure and dynamics, which are revealed in extraordinary detail by modern diffraction and spectroscopy experiments. 
The results of these experiments are thus a natural place to begin building models of entropy. 
Unfortunately, however, these data, particularly when taken individually, may give an ambiguous or even misleading picture of contributions to a material's entropy.
The purpose of this Perspective is to review pitfalls in interpreting crystallographic and spectroscopic data collected from disordered crystalline materials, with particular reference to modelling entropy directly from these data.

In the following two subsections, we outline the two main contributions to entropy we will discuss -- vibrational and configurational entropy (Section~\ref{sec:decomposing}) -- and present a simple toy model of a phase transition that illustrates the ways in which these forms of entropy can occur (Section~\ref{sec:toy}). We then consider the effects of each of these contributions, first on crystallography (Section~\ref{sec:cryst}), then on neutron spectroscopy (Section~\ref{sec:spec}).

\subsection{Decomposing contributions to entropy}\label{sec:decomposing}
Of course, the problem of relating entropy to structure is relevant far beyond the specific application to barocalorics.
In a sense, the entire discipline of materials chemistry is founded on the understanding that the bulk thermodynamics that govern a substance's stability and reactivity are determined by its atomic-level structure and dynamics. 
The first step in understanding this relationship is to calculate the internal energy associated with a particular atomic configuration, the realm of quantum chemistry. 
But when working at constant temperature, as in most practical cases, it is the free energy that must be minimised; in this case, the entropy of an atomic configuration also becomes important and indeed may dominate the free energy difference between competing phases.\cite{nyman_static_2015, butler_microscopic_2016, mendels_searching_2018} Accurate determination of the entropy of a given material structure is thus just as important as the internal energy for rationalising known crystal structures, crystal structure prediction and crystal engineering and design.\cite{reilly_report_2016} 


Both energy and entropy are extensive state functions that can therefore be viewed as the sum of contributions from different atoms or physical origins. For instance, the cohesive energy of a molecular salt might include contributions from dispersion forces, hydrogen-bonded pairs, and the Coulomb force; these contributions can be visualised and understood in terms of ``energy frameworks''.\cite{turner_energy_2015, mackenzie_crystalexplorer_2017} 
The entropy can in principle be decomposed in a similar way.\cite{butler_organised_2016} 

Here, we will consider the two main effects in non-magnetic, crystalline materials: vibrational and configurational contributions. This neglects, for instance, the electronic entropy (relevant to systems with spin degrees of freedom)\cite{vallone_giant_2019} and the entropy of free rotation (most relevant to gas-phase molecules).\cite{garcia-ben_discovery_2022} Broadly speaking, the vibrational entropy describes a system's movement within a single basin of the energy hypersurface, while the configurational entropy describes its choice between multiple basins. 
We will expand on this distinction in the following section, but there are three important points to make at the outset.

First, different sorts of entropy require different mathematical formalisms.\cite{fultz_vibrational_2010} We must correctly identify which is relevant to a given material in order to perform these calculations correctly. Specifically, the configurational entropy is straightforwardly given by the Boltzmann formula
\begin{equation}
    \label{eq:Sconf}
    S = k\ln n,
\end{equation}
where $k$ is the Boltzmann constant and $n$ the number of equivalent configurations a structure can occupy.\footnote{In principle, this formula applies generally, with $n$ being the number of microstates accessible to a system in a phase space including both positions and momenta. Here we will apply it directly only in the case where $n$ represents different, equivalent, positions, for instance in crystallographic disorder models. In the context of barocaloric materials, this is equivalent to assuming that the local shape of the potential well is similar enough in the high- and low-entropy phases that vibrational contributions effectively cancel.}
The vibrational entropy, by contrast, arises from the standard expression for a harmonic oscillator
\begin{equation}
    \label{eq:Svib}
    S = k\big((n+1)\ln(n+1) - n\ln n\big),
\end{equation}
where now $n(E,T)$ is the \emph{number of phonons} in a particular mode at a given temperature. In practice, many modes will be active and this is more typically used as an integral over the phonon density of states $g(E)$:
\begin{equation}
    \label{eq:Svibg}
    S = 3k\int_0^\infty g(E)\big((n(E)+1)\ln(n(E)+1) - n(E)\ln n(E)\big)\,\mathrm{d}E.
\end{equation}

Second, this description immediately illustrates that there is no clear division between these types of entropy; our perspective may change, for instance, depending on the barrier height between basins (or, equivalently, the temperature).  

Third and finally, this additive property is a double-edged sword, in the sense that these calculations reduce complex configurations of atoms to single scalar numbers: it is thus easy for errors in different components to accumulate or, conversely, to cancel fortuitously. As a result, it is difficult to judge the accuracy of an entropy calculation, or to troubleshoot it, simply by comparison with the experimentally determined value.


\subsection{A toy model}\label{sec:toy}

\begin{figure}[tbhp]
  \centering
  \includegraphics[width=0.45\textwidth]{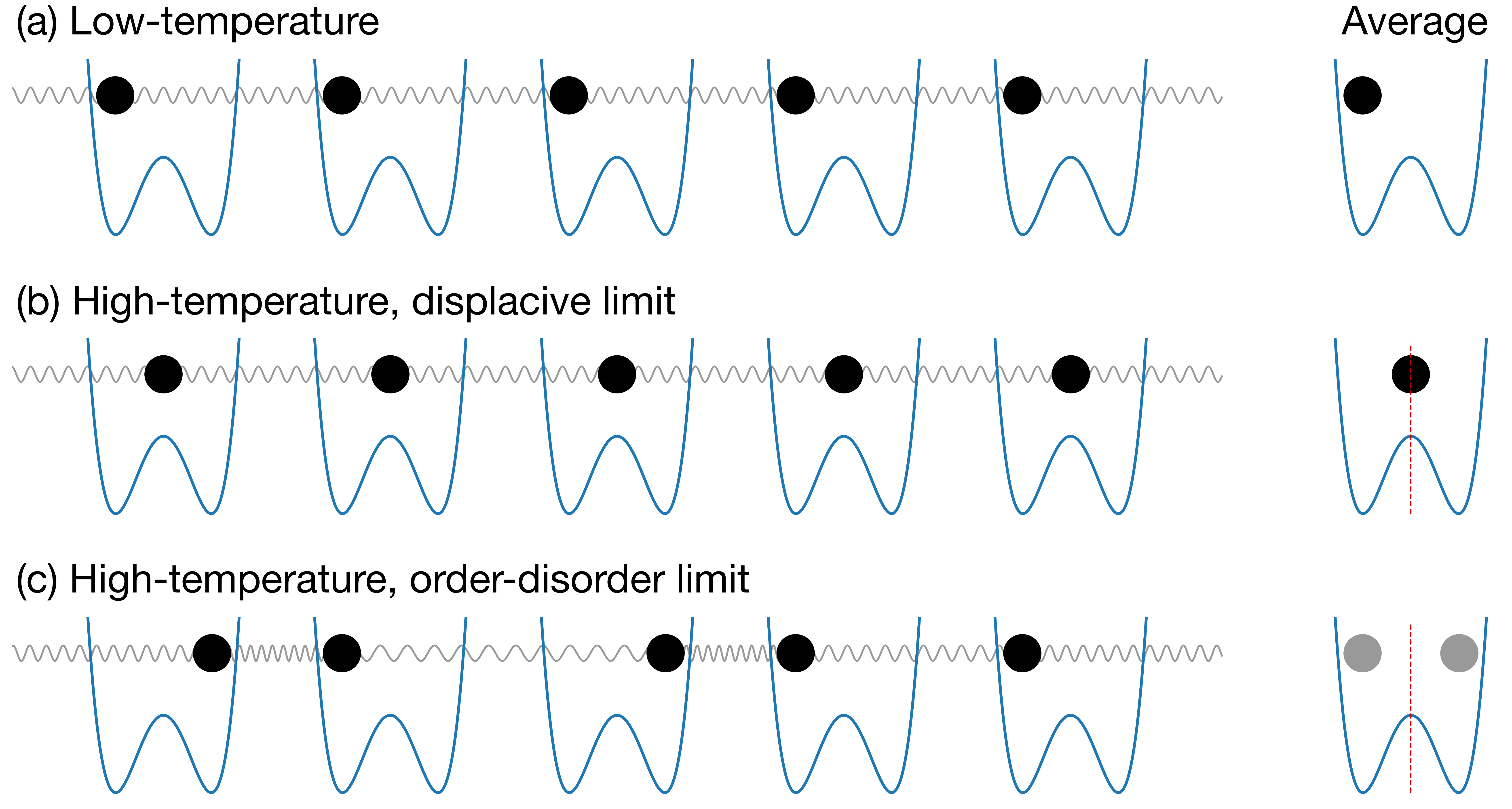}
  \caption{(a) A toy model of a phase transition in which each particle moves in a quartic double-well potential, and is connected to its nearest neighbours by a harmonic spring. Depending on the relative strengths of the double-well and spring potentials, this can give a spectrum of high-temperature phases, ranging from (b) displacive to (c) order-disorder. The average structure is shown on the right-hand side. The new symmetry element introduced in the high-temperature phase is the mirror plane indicated here by a dotted red line.}
  \label{fig:pt_types}
\end{figure}

Since barocaloric materials require a phase transition between a high- and a low-entropy phase, we consider here a simple model of the entropy in such phase transitions.
We assume that the space-group symmetries of the two phases are related by a group-subgroup relationship: in other words, that moving from the low- to the high-temperature phase introduces a specific new symmetry element to the crystallographic structure. In Fig.~\ref{fig:pt_types}, this symmetry element is the mirror plane that maps the left half of each site to the right. We can distinguish two different ways in which this symmetry change might be achieved. In a \emph{displacive} phase transition (Fig.~\ref{fig:pt_types}b), each site locally adopts this higher symmetry. By contrast, in an \emph{order-disorder} phase transition (Fig.~\ref{fig:pt_types}c), each atomic site retains the same local symmetry as in the low-symmetry phase; the higher symmetry emerges only on taking the average of every site in the crystal. 

These two types of phase transition map neatly onto the two contributions to entropy discussed in Section~\ref{sec:decomposing}. In the displacive case, each atom moves freely between the left and right sides of its site, and the entropy difference between the phases is exclusively vibrational. In the order-disorder case, each atom is confined to the well on a single side, just as it was in the low-temperature phase, and the entropy difference is exclusively configurational.


As a model of the physical origin of such behaviour, anticipated in Figure~\ref{fig:pt_types}, suppose that each atom sits in a local double-well potential, and is connected to its nearest neighbours by harmonic springs.\cite{bruce_structural_1980-ii}
The total potential energy is thus
\begin{equation} 
    U = \sum_i\left(-\tfrac12\kappa_2u_i^2 + \tfrac14\kappa_4u_i^4\right) + \tfrac12J\sum_{i, j}^\mathrm{n.n.}(u_i - u_j)^2
\end{equation}
where the $u_i$ represent the displacements of each atom from the centre of its site; $\kappa_2$ and $\kappa_4$ characterise the double-well potential; $J$ characterises the inter-site ``spring'' interaction; and the second sum is to be taken over all pairs of nearest neighbours (``n.n.'') $i$ and $j$.

In this model, the competition between the site potentials and the nearest-neighbour interactions is encapsulated in the parameter $s \equiv \kappa_2/8J$. If $s\ll1$, the nearest-neighbour interactions dominate over the site potentials; the phase transition is displacive; and the entropy change is entirely vibrational. If $s \gg1$, the site potentials dominate; the phase transition is order-disorder; and the entropy is purely configurational.\cite{dove_theory_1997} Vitally, though, this model demonstrates that there is a continuum between these two limits, so that a real phase transition may combine features of both sorts.



Despite its simplicity, this model is difficult to analyse in full generality. 
Two ways of simplifying it are to assume that either the \emph{sites} or the vibrational \emph{modes} are independent of one another.\cite{bruce_structural_1980-ii}
The first of these, the independent site approximation, gives rise to a mean-field, Landau-like theory.
The second gives a theory in terms of harmonic phonon modes. 
This is not as restrictive as it may at first seem, since it is possible to use quasiharmonic approaches with renormalised mode frequencies that take account of anharmonic behaviour -- although this approach depends on being able to select the correct local minimum to explore.\cite{abraham_adding_2019}
We will use this second approach in much of the following analysis.

Alternatively, this simplification can be avoided entirely by using molecular dynamics simulation to directly sample the relevant region of the potential energy surface.\cite{dybeck_capturing_2017} This is an accurate and very general approach, although it may be expensive depending on the potentials used. In particular, it avoids the need to partition entropy into vibrational and configurational contributions, which could be an advantage particularly for intermediate cases, but which at the same time makes the resulting entropy more difficult to interrogate for individual structural contributions.

\section{Crystallography}\label{sec:cryst}



The intensities of Bragg peaks depend on the space and time average of the contents of the crystallographic unit cell. Specifically, they are sensitive to the scattering density within this cell. In theory, this scattering density could be represented in terms of a wide range of possible basis functions. In practice, standard crystallographic models represent it as the sum of ellipsoids, each representing a single atom, as a way of achieving a good fit with relatively few refined parameters.
The dimensions of these ellipsoids are the atomic displacement parameters (ADPs); these were earlier called ``thermal parameters,'' but the change of name reflects that these represent displacement from the average site for any reason, including static disorder as well as thermal motion. Inasmuch as they do represent thermal vibrations, however, the choice to model each atom as an ellipsoid is equivalent to taking the (quasi)harmonic approximation.\cite{willis_thermal_1975} 

We note in passing that an important research theme in crystallography aims to go beyond this ``independent atom model,'' using anisotropic atomic form factors that can be refined against very high-quality data, determined by quantum chemistry calculations, and/or taken from standard reference databases.\cite{dittrich_generalized_2013, jha_multipolar_2022} These methods, however, are not generally appropriate for the highly disordered phases common in barocaloric materials, which occur at relatively high temperatures and for which far fewer independent reflexions (data points) are generally available. 

There are well-established methods to determine both the vibrational and configurational entropy from such crystallographic models.
The vibrational entropy can be calculated either directly from the temperature dependence of the ADPs (normal coordinate analysis)\cite{burgi_dynamics_2000, saito_entropic_2005} or by refining instead the population of vibrational modes determined from periodic DFT calculations.\cite{kofoed_x-ray_2019}
The configurational entropy, on the other hand, has been related to the Shannon information that is encoded in a crystal structure where atoms are ``labelled'' by their crystallographic orbits.\cite{krivovichev_structural_2016, hornfeck_extension_2020} This can now be calculated routinely by the program crystIT.\cite{kaussler_crystit_2021}

In crystallographically ordered materials, the representation of the scattering density in terms of atomic ellipsoids is essentially unique.
In the context of disordered materials, however, several equally valid representations of the \emph{same} scattering density may be possible; an unwary scientist may take the differences between these representations more seriously than the underlying data warrant. 

\begin{figure}[hbt]
  \centering
  \includegraphics[width=0.45\textwidth]{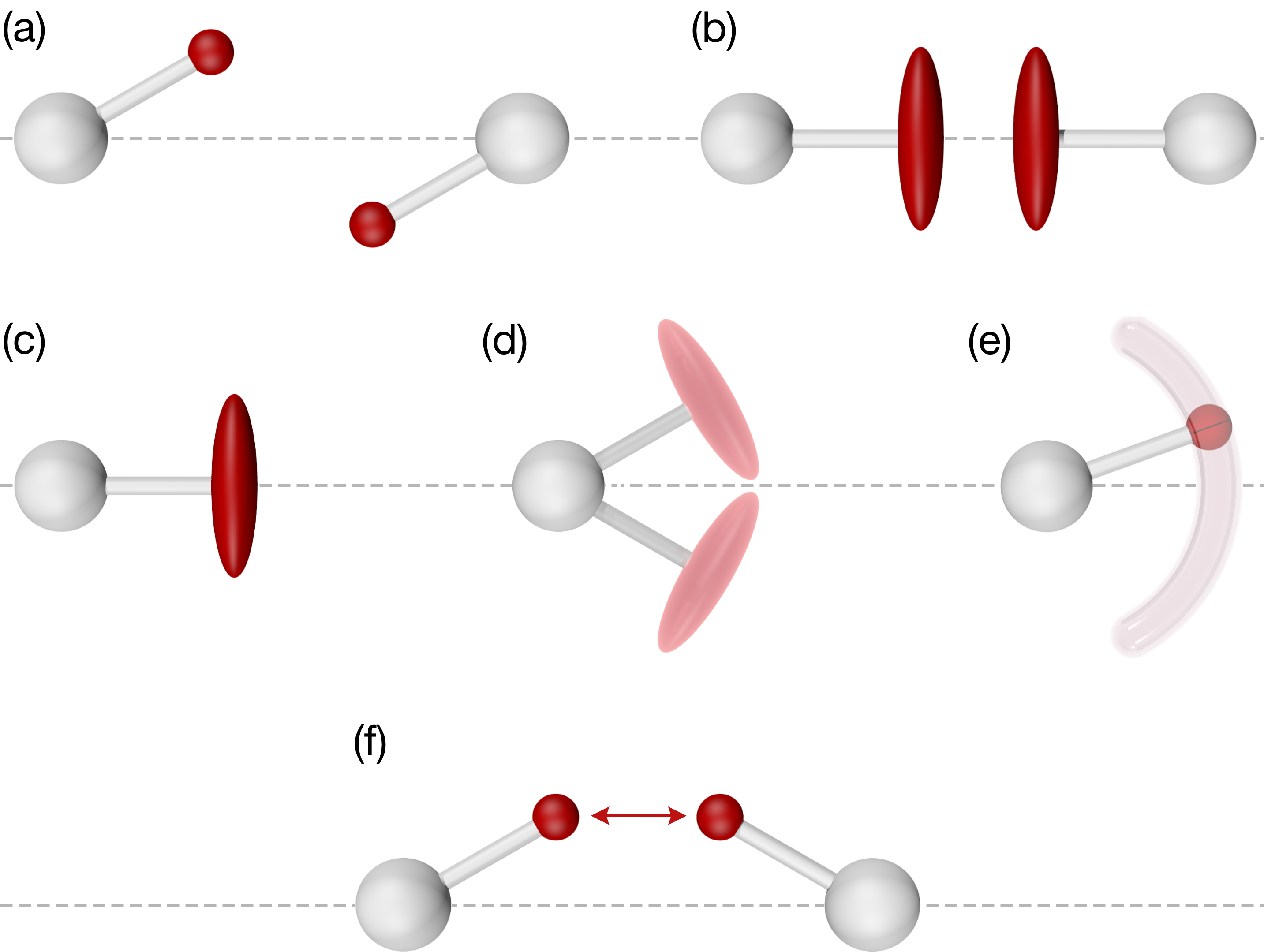}
  \caption{An example of a typical phase transition, in which the high-symmetry phase introduces an additional mirror plane (dotted line) compared to the low-symmetry phase. Shown first are the initial structural solutions of (a) the low-temperature phase and (b) the high-temperature phase. In the high-temperature phase, both (c) one-site and (d) two-side disorder models may give reasonable fits, while (e) the true distribution of scattering density may not be ellipsoidal. (f) Close contact between particular partially occupied sites may reduce the configurational entropy.}
  \label{fig:disorder_types}
\end{figure}

As an example, consider a hypothetical phase transition that creates a crystallographic mirror plane. In the low-symmetry phase (Fig.~\ref{fig:disorder_types}a), there are clear peaks of scattering density on one or other side of this plane, revealing the locations of atoms; while in the high-symmetry phase, the scattering density is smeared symmetrically across the plane (Fig.~\ref{fig:disorder_types}b).
%
There are at least two plausible crystallographic models of the disordered phase: we might choose to use a single, fully occupied site \emph{on} the plane (Fig.~\ref{fig:disorder_types}c) or two half-occupied sites on either side of the plane, related by the mirror symmetry (Fig.~\ref{fig:disorder_types}d).

These two models appear to correspond neatly to the two extremes of entropy discussed in Section 1: the single-site model to a displacive phase transition in which the entropy is mainly vibrational; the split-site model to an order-disorder phase transition in which the entropy is mainly configurational. For this reason, it is tempting to claim that these two types of phase transition can be distinguished by trying both models and choosing the one with the lower residual ($R$-factor).
It would, however, be a serious mistake to decide between these two cases on the basis of the crystallographic model alone. The split-site model is likely to give an better fit, regardless of the true distribution of atomic positions, for at least two reasons.

First, the split-site model, where the site is on a general position, has nine parameters: three to define the position vector, six to define the ellipsoid, which is represented by a rank-two symmetrical tensor. On the other hand, the single-site model, where the site is on the mirror plane, has only two position parameters (since it is free to move only on the 2D plane) and only four displacement parameters (which can be thought of as the length, width and breadth of the ellipsoid plus a rotation perpendicular to the mirror plane), for a total of only six parameters. The split-site model therefore has three more degrees of freedom, which are available to absorb noise even if they are not required by the true scattering density; it would be expected to give a lower $R$-value for this reason alone. At the least, therefore, a statistical correction for this effect is required to meaningfully compare these models.\cite{hamilton_significance_1965}

Second, if the atom vibrates freely across the mirror plane, the length of the bond with its neighbour will remain roughly constant, so the distribution of scattering density is likely to be curved or ``banana-shaped'' (Fig.~\ref{fig:disorder_types}e). If this curvature is pronounced, then it will be better fitted by two ellipsoids at an angle, as in the split-site model (Fig.~\ref{fig:disorder_types}d), than a single ellipsoid perpendicular to the plane, as in the single-site model (Fig.~\ref{fig:disorder_types}c), even if the distribution of scattering density is indeed centered on the mirror plane! In principle, this could be accomodated by describing the scattering density in terms of a more sophisticated basis set, but this is unlikely to be practical for the reasons given above.

Suppose, though, that we have distinguished reliably between these two cases, concluding that the entropy is indeed primarily configurational. Even now, we cannot directly calculate the system's entropy.

As noted above, the relevant formula for configurational entropy is the Boltzmann equation \eqref{eq:Sconf}. If the two split sites are independent, each will therefore contribute $R\ln 2$ to the entropy, for a total of $R\ln 4$. On the other hand, if (say) steric hindrance prevents the two atoms from being on the same side of the mirror plane at the same time, as suggested by the low-temperature structure (Fig.~\ref{fig:disorder_types}f) -- or indeed, in a different situation, if there is an attractive force so that the two atoms \emph{prefer} to be on the same side of the mirror plane -- there will only be two possibilities and the entropy will be only $R\ln 2$.\cite{saito_configurational_2005}

Naturally, a continuum is possible between these two cases: that is, two adjacent atoms may be sterically unfavourable and therefore downweighted without actually being impossible. In other words, the fact that all of the crystallographic split sites in this model have the same average occupancy of $1/2$ does \emph{not} mean that all of the possible local configurations have equal probability.

In summary, even a well-fitting crystallographic model of a disordered phase may be ambiguous as to whether the entropy is primarily vibrational or configurational. We will now see that exactly the same is true of vibrational spectroscopy.


\section{Spectroscopy}\label{sec:spec}
A perfect, static crystal has zero entropy, so that entropic contributions must be inferred indirectly from crystallographic data in the ways discussed above. On the other hand, vibrational spectroscopy inherently reveals entropy due to this vibrational motion, which indeed can be calculated as an integral over the phonon density of states \eqref{eq:Svibg}. In addition, however, further signatures of disorder are often observed in inelastic neutron spectra, which may reveal additional contributions to the entropy. Specifically, in contrast
with the sharp, detailed phonon spectra typically produced by \emph{ab~initio} calculations, those measured experimentally 
display varying degrees of broadening. 

Broadening arising from the instrumental resolution can be minimised by good instrument and experimental design, but cannot be removed entirely. It is therefore essential to have a good understanding of the resolution to determine whether the observed broadening has an additional physical origin. Historically, the understanding of the resolution ellipsoid on triple axis spectrometers has been very well understood, allowing an optimised data collection strategy and deconvolution from the data. For time-of-flight direct geometry spectrometer measurements, which yield a survey of excitations over a much larger region of reciprocal space, this has been more complicated, but with the development of software like Euphonic,\cite{Fair} the full $Q$-$E$ resolution (that is, with respect to both momentum and energy transfer) can now be included in a complete data analysis.

Once the instrumental resolution has been taken into account, any further broadening must have physical significance for the system being studied. As before, it is convenient to divide the potential causes into two main cases, depending on whether the disorder is dynamic or static on the timescale set by the phonon frequency. 
The static case corresponds more or less directly to the case of configurational entropy discussed above, while the dynamic case will alter the value of the vibrational entropy.
As well as this characteristic time scale, phonons also have a characteristic length scale set by their wavelength. Thus long-wavelength, low-frequency vibrations are likely to sample a range of whatever parameter characterises the disorder (for instance, different site occupancies or molecular orientations), while short-wavelength, high-frequency vibrations may instead reflect a particular local configuration of atoms.
Once again, there is no clear dividing line between these two origins. Indeed, a common and technologically relevant intermediate situation involves orientational disorder of molecules in crystals, which may be anywhere from frozen to almost unhindered depending on the phase and temperature, and which will couple to the translations involved in vibrational motion.\cite{lynden-bell_translation-rotation_1994} Furthermore, some forms of disorder may fall into \emph{both} categories: for instance, isotopic disorder is clearly configurational (static disorder), but also, because of the mass difference between isotopes, disrupts harmonic phonons (dynamic disorder).\cite{maradudin_theory_1993}
Nonetheless, the static and dynamic extreme cases can be thought of in different ways and contribute differently to the entropy, and it is worthwhile to distinguish them by a thorough analysis of the experimental data. 

We consider first the case of dynamic disorder. 
Phonons are quantum-mechanical quasiparticles that arise from quantising the vibrational normal modes of a crystal. Typically, atomic displacements in these modes are small, allowing the potential of the lattice to be written in terms of a Taylor expansion about the equilibrium position. The harmonic approximation corresponds to truncating the expansion at the second order term, valid for many small displacements, resulting in a harmonic oscillator.
As noted above, this approximation is equivalent to assuming that the vibrational modes do not interact, and gives excitations with infinite lifetimes and sharp peaks. On the other hand, in systems where the potential is appreciably anharmonic, there will be phonon-phonon interactions which will reduce the phonon lifetime, resulting in broadening in energy. Using lowest-order perturbation theory, it can be shown that this damping is linearly proportional to temperature. 
Furthermore, in the anharmonic case, the phonon frequency is no longer independent of the mode amplitude, so that a frequency shift with temperature is also expected.
Such phonon anharmonicity is significant for many important condensed matter phenomena, including negative thermal expansion, ultralow thermal conductivity, high-temperature superconductivity, and soft mode phase transitions. An in-depth understanding of the microscopic mechanism is thus vital for many applications. 

\begin{figure}[hbt]
  \centering
  \includegraphics[width=0.45\textwidth]{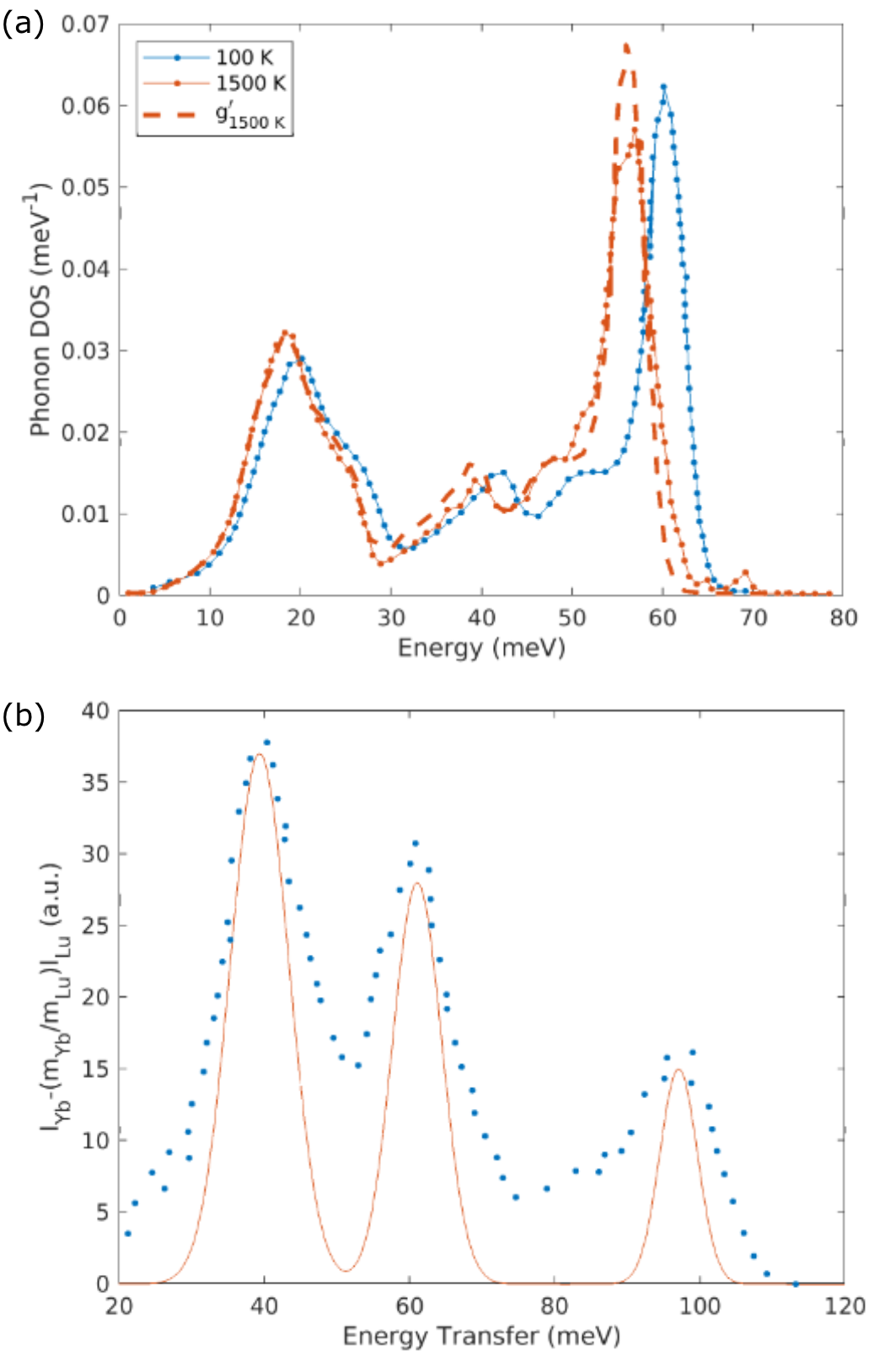}
  \caption{Two examples of broadening of INS peaks due to disorder. (a) In silicon, the peak at 60~meV in the phonon density of states at low temperature (blue) shifts in frequency on heating (red), and is furthermore broadened with respect to the rescaled low-temperature data (red dashed line). Both of these effects are \emph{vibrational}, due to phonon anharmonicity. Data taken from ref.~\citenum{Kim}. (b) In \ce{YbMgGaO4}, crystalline electric field excitations of the \ce{Yb^3+} ion (integrated over $4 \leq Q \leq 6$~\si{\angstrom^{-1}}, blue) are broader than the instrumental resolution function (orange). This time the effects are due to \emph{configurational} disorder due to site mixing between \ce{Mg^2+} and \ce{Ga^3+} ions, although the broadening is superficially similar to (a). Data taken from ref.~\citenum{Li}.}
  \label{fig:INS_broadening}
\end{figure}

Phonon anharmonicity has been explored extensively using inelastic neutron scattering, through the measurement of phonon dispersion curves, phonon lifetimes and phonon densities of states. With no claim to comprehensiveness, we give here a few recent examples. In the thermoelectric PbTe, INS reveals strong broadening and softening with increasing temperature, indicative of substantial anharmonicity. This is amplified by nesting between transverse acoustic and optical modes, which enables more three-phonon scattering channels.\cite{Delaire} In the prototypical hybrid perovskite semiconductor MAPI (\ce{CH3NH3PbI3}), triple-axis measurements show that the acoustic phonon lifetimes are 50--500 times shorter than those in conventional semiconductors, due both to strong phonon-phonon interactions and to coupling with rotations of the methylammonium ions. This could have significant implications for their application due to the impact on thermal conductivity affecting hot carrier cooling.\cite{Gold-Parker} The effects of anharmonicity can also be seen in phonon density of states measurements obtained on polycrystalline samples. In the case of silicon, a systematic softening and broadening are seen in the spectra with increasing temperature, where the shifts to lower energies at high temperatures are too large to be accounted by simple quasiharmonic behaviour, while both the broadening and $80\%$ of the softening can be accounted for by phonon anharmonicity (Fig.~\ref{fig:INS_broadening}(a)).\cite{Kim} Once identified, anharmonicity can be incorporated in entropy calculations by mapping the energy of the system as a function of the relevant mode amplitudes, and hence calculating the partition function from which the free energy or entropy can be derived.\cite{skelton_anharmonicity_2016}

We now move to the second case, broadening of vibrational peaks caused by static disorder. 
One simple example is the broadening of crystal field levels well beyond the instrumental resolution in the putative quantum spin liquid \ce{YbMgGaO4}. This was ascribed to a random crystalline electric field arising from site mixing between \ce{Mg^2+} and \ce{Ga^3+} affecting the local coordination of the \ce{Yb3+} ions, suggesting that it was inherent structural disorder responsible for the spin-liquid physics (Fig.~\ref{fig:INS_broadening}(b)).\cite{Li}

Well-studied early examples of static disorder include the asymmetrical linear molecule \ce{N2O},\cite{lemaistre_raman_1988} which freezes into one of two orientations in the solid state, with little short-range order or subsequent reorientation. It has long been recognised that molecular dynamics simulations are an ideal means to study such materials,\cite{cardini_molecular_1990, cardini_molecular_1991} but the potential scope of such simulations has been broadened tremendously both by increases in computing power and by recent developments in interatomic potentials.\cite{unke_machine_2021}
This in turn makes it possible to examine far more complex functional materials with inherent disorder using INS, and to interpret the resulting data in terms of specific disorder models.

The effects of static disorder on the vibrational spectra of crystalline materials have been modelled largely in one of two limits of computational difficulty. First, the virtual crystal approximation looks at the average effect of disorder on phonon dispersion, treating the phonon broadening as a perturbation. This typically works well only for low-frequency phonon modes and for low levels of disorder. Alternatively, supercell-based methods allow disorder to be accounted for explicitly, but are restricted by the computational cost of performing phonon (or molecular dynamics) calculations on large simulation cells. In this second case, the cost may be ameliorated by using a cheaper Hamiltonian. For instance, while density-functional theory is practically restricted to 
relatively small systems, perhaps $\sim100$--$1000$ atoms, suitably parametrised empirical potentials are far less expensive and can therefore be used to model larger systems.
In either case, ``unfolding'' the supercell phonon modes onto the Brillouin zone of the original crystal -- an approach known as
supercell lattice dynamics (SCLD)\cite{Overy1,Overy2} -- will then yield dispersion curves with static disorder accounted for.

We recently validated this approach against experimental INS data. Adamantane is a barocaloric plastic crystal that undergoes a phase transition at $T=208$~K to an orientationally disordered phase,\cite{Beake} in which significant broadening and softening of the acoustic modes is observed. These modes are responsible for most of the vibrational entropy in adamantane, making it important to understand whether these effects arise due to anharmonicity or disorder. The SCLD calculations reproduce a substantial amount of the phonon broadening seen, indicating that it can be attributed in large part to orientational disorder, without the need to further take into account rotation-translation coupling.\cite{Meijer}

As simulation methods develop further, it will be possible to explore the phonons in systems with multiple forms of disorder. This has already been demonstrated in the case of the high-entropy alloy FeCoCrMnNi, which as a random solid solution is disordered with respect to atomic size, mass and force constants, but still displays clear features in the phonon spectra. Phonon broadening is observed with an anisotropic $Q$-dependence, while no significant temperature dependence is observed for the phonon frequency or spectral shape, leading to the exclusion of anharmonicity or magnetic fluctuations being the dominant mechanism. Using first-principles calculations combined with the itinerant coherent potential approximation and supercell phonon-unfolding methods,\cite{Mu} it could be demonstrated that force-constant disorder is predominantly responsible for the phonon lifetime reduction.\cite{Turner} 

We note finally that the phonon formalism applies specifically to crystalline materials, even if disordered; for amorphous materials, other descriptions of the excitations will be more suitable.\cite{allen_diffusons_1999} Indeed, these may be preferable even with impurity concentrations as low as a few percent.\cite{seyf_rethinking_2017} Such alternative models, are, however, beyond the scope of this brief perspective.

\section{Conclusions}

Structural disorder reveals itself with well-known signatures in both crystallographic and spectroscopic data: typically, crystallographic models show large ADPs and atom sites that ``may be split'', while excitation peaks are broadened. Attempting to determine a material's entropy directly from such data, however, can be dangerous, specifically because configurational and vibrational components are easy to confuse but contribute in very different ways to the total entropy. 

Of course, diffraction and spectroscopy remain vital tools in characterising disordered phases, including those of barocaloric materials. Certainly, any convincing mechanism for a giant entropy change must be consistent with both crystallography and spectroscopy. One way to be more confident in an analysis of contributions to the entropy is simply to combine crystallographic and spectroscopic data. Another is to cross-reference against additional techniques, for instance diffuse scattering or solid-state NMR. In general, to avoid confusion between different types of entropy, great care should be taken to predict the experimentally observable consequences of a given structural model, and to check that these are in fact distinguishable from those predicted by alternative plausible models.

We conclude with a brief look to the future. The problem of deriving entropy from a structural model is in practice less routinely solved than the corresponding problem for energy. There is, however, no fundamental reason why the crystal engineering community could not target a high-entropy structure in much the same way, and with the same accuracy, as is now done to design low-energy states. This would be an extremely important step towards designing functional materials including dielectrics and ionic conductors as well, of course, as calorics.

\section*{Acknowledgements}
We thank Prof. Martin T. Dove for very helpful discussions.

\providecommand*{\mcitethebibliography}{\thebibliography}
\csname @ifundefined\endcsname{endmcitethebibliography}
{\let\endmcitethebibliography\endthebibliography}{}


\end{document}